\documentstyle[twoside, epsf]{article}

\input ibvs2.sty

\newcommand{\pn}{\phantom{0}}

\begin{document}

\IBVShead{5xxx}{00 Month 2009}

\IBVStitle{Updated spin ephemeris for the cataclysmic variable EX Hydrae}

\IBVSauth{MAUCHE, C.W.$^1$; BRICKHOUSE, N.S.$^2$; HOOGERWERF, R.$^3$;
           Luna, G.J.M.$^2$; Mukai,~K.$^4$; STERKEN,~C.$^5$}

\IBVSinst{Lawrence Livermore National Laboratory,
           L-473, 7000 East Avenue, Livermore, CA 94550, USA,
           e-mail: mauche@cygnus.llnl.gov}
\IBVSinst{Harvard-Smithsonian Center for Astrophysics,
           60 Garden Street, MS-15, Cambridge, MA 02138, USA,
           e-mail: nbrickhouse@cfa.harvard.edu, gluna@cfa.harvard.edu}
\IBVSinst{Interactive Supercomputing, Inc.,
           135 Beaver Street, Waltam, MA 02452, USA,
           e-mail: hoogerw@pobox.com}
\IBVSinst{NASA GSFC, Code 662, Greenbelt, MD 20771, USA,
           e-mail: mukai@milkyway.gsfc.nasa.gov}
\IBVSinst{Vrije Universiteit Brussel, Pleinlaan 2, B-1050 Brussels, Belgium,
           e-mail: csterken@vub.ac.be}

% V* EX Hya  -- Cataclysmic Var. DQ Her type
% FK5 coord. (ep=2000 eq=2000) : 12 52 24.40 -29 14 56.7
% aka 4U 1249-28, 3A 1250-289

\SIMBADobjAliasJ2000{EX Hya}{4U 1249-28}{12 52 24.40 -29 14 56.7}
%\IBVSobj{EX Hya}
\IBVStyp{UGSU+E}
\IBVSkey{photometry}

\IBVSabs{Historical optical data are combined with more recent}
\IBVSabs{optical, extreme ultraviolet, and X-ray data to update}
\IBVSabs{the spin ephemeris of the cataclysmic variable EX~Hya.}

\begintext

Recent satellite observations demonstrate that the phase of maximum flux of the
67 min spin modulation of the white dwarf in the cataclysmic variable EX~Hya is 
drifting away from the optical quadratic ephemeris of Hellier \& Sproats (1992, 
hereafter HS92). Relative to that ephemeris, the peak of the spin-phase extreme 
ultraviolet (EUV) flux modulation measured with the {\it Extreme Ultraviolet 
Explorer\/} ({\it EUVE\/}) was $\phi_{67}=0.040\pm 0.002$ in 1994 May (Mauche 
1999) and $\phi_{67}=0.115\pm 0.001$ in 2000 May (Belle et al.\ 2002). 
Similarly, the peak of the spin-phase X-ray flux modulation measured with the 
{\it Chandra X-ray Observatory\/} was $\phi_{67}\approx 0.1$ in 2000 May 
(Hoogerwerf, Brickhouse, \& Mauche 2004) and $\phi_{67}\approx 0.2$ in 2007 May 
(Luna, Brickhouse, \& Mauche 2008). Because the discrepancy between the observed 
$O$ and calculated $C$ phases of the spin-phase flux modulation of EX~Hya is now 
approaching a significant fraction of a spin cycle, we have undertaken the task 
of updating the ephemeris.

Toward that end, we have have combined the optical data of
Vogt, Krzeminski, \& Sterken (1980, hereafter VKS80),
Gilliland (1982),
Sterken et al.\ (1983),
Hill \& Watson (1984),
Jablonski \& Busko (1985),
Bond \& Freeth (1988),
HS92,
Walker \& Allen (2000), and
Belle et al.\ (2005)
with the optical, EUV, and X-ray data listed in Table~1. The first set of 
optical data in Table~1 was obtained by CS at the European Southern 
Observatory, La Silla, Chile using the Danish 1.5-m telescope and the DFOSC CCD 
camera. Differential $V$-band magnitudes were obtained by aperture photometry 
extracted from flat-fielded and bias-corrected CCD frames. The second set of 
optical data in Table~1 was obtained by Beuermann \& Reinsch (2008, hereafter 
BR08) and is included here to clear up an ambiguity in the units of the timings
in their Table 3, which are labeled as HJD, described as BJD, and treated as
BJD(TT), whereas they are in fact BJD(UT); this change affects all the $O-C$
values in their table. Other than the {\it EXOSAT\/}, {\it Ginga\/}, and BR08
data, which have been taken from the given references, all other times of spin
maximum in the table have been derived by us from the various datasets. In the
processes, we have corrected an error in the (spin {\it and\/} orbit) phases
of the {\it ASCA\/} data published by Ishida, Mukai, \& Osborne (1994) and the
{\it RXTE\/} data published by Mukai et al.\ (1998). We note that our result
for the second {\it EUVE\/} observation agrees within the errors with the 
result derived independently by Belle et al.\ (2002). Table~1 lists the observed 
times of spin maximum in Barycentric Julian Date, the corresponding cycle number 
$E$ derived from the HS92 quadratic ephemeris, and the $O-C$ residuals in days 
relative to the VKS80 linear ephemeris, the HS92 quadratic ephemeris, and our 
cubic ephemeris (eqn.~1).

The task of combining optical, EUV, and X-ray data into a single ephemeris
presents a number of challenges. First, the published times of optical flux 
maximum typically do not include error estimates. Second, the times of flux 
maximum are typically determined in different manners in the optical and
higher-energy wavebands. In the optical, the {\it times\/} of the flux maxima
are typically estimated directly from the light curves, whereas in the EUV
and X-ray wavebands, where the event rates are often fairly low, the events
are typically phase-folded to produce a mean light curve, from which the
{\it phase offset\/} relative to the assumed ephemeris is calculated from an 
analytic (typically, sine) fit to the mean light curve. From this, the
effective time of flux maximum is derived, typically referenced to the start
or mid-point of the observation. This approach is capable of producing very
high signal-to-noise ratio light curves and hence error values on the fit
parameters, particularly the times of flux maxima, that are formally very
small.

Given these complications, we have taken a multi-step approach to calculate a
revised spin ephemeris for EX~Hya. First, we fit the optical data to a quadratic
ephemeris without weights, producing the ephemeris constants listed in the first
entry of Table~2. The standard deviation of this fit is 0.00360 days or 0.077
cycles (which, if used as a uniform error on the data, produces the same fit
with a reduced $\chi^2 =1$). Second, we fit the EUV and X-ray data to a 
quadratic ephemeris accounting for the errors listed in Table~1, producing the 
ephemeris constants listed in the second entry of Table~2. The two results, 
optical on one hand and EUV and X-ray on the other, are consistent within the 
errors and are as well close to (but different from) the optical quadratic 
ephemeris constants of HS92. Next, we fit the combined data sets, using 0.00360 
days for the error on the optical data and the errors listed in Table~1 for the 
errors on the EUV and X-ray data, producing the ephemeris constants listed in 
the third entry of Table~2. The ephemeris constants are now significantly 
different from those of the previous fits, although it is apparent that the fit 
is not ideal ($\chi^2$ per degree of freedom (dof) $=651.2/431 =1.51$), in part 
because the ephemeris rolls over too rapidly at early times. To remedy this 
deficiency, we fit the combined data sets to a cubic ephemeris, producing the 
ephemeris constants listed in the fourth entry of Table~2. The fit is now 
somewhat improved ($\chi^2/{\rm dof} =638.5/430 = 1.48$), the fit parameters
are  closer to those of the earlier quadratic fits, the ephemeris is close to
that of HS92 through 1991 January (230{,}000 cycles; Fig.\ 1{\it a\/}), and it 
reproduces well all of the available EUV and X-ray data (Fig.\ 1{\it c\/}). 
Finally, by setting a lower limit of 0.02 cycles or 0.00093 days on the size of 
the timing errors on the EUV and X-ray data, the reduced $\chi^2$ of the fit
is  reduced to a very reasonable $\chi^2/{\rm dof} =471.0/430 = 1.10$. Based
on  these results, we recommend that the following cubic ephemeris be used for 
recent past and future timings of the flux maxima of the spin modulation of the 
white dwarf in EX~Hya:

\begin{equation}
T_{\rm max} =
   2437699.8917(6) + 0.046546484(9) \, E -  7.3(4) \times 10^{-13}\, E^2
   + 2.2(6) \times 10^{-19}\, E^3.
\end{equation}

\clearpage

\noindent{\bf Acknowledgements:}
The ESO La Silla optical data used in the work were obtained with the Danish 
1.5-m telescope, which is operated by the Astronomical Observatory, Niels Bohr
Institute, Copenhagen University, Denmark. We thank K.\ Beuermann for clearing
up the ambiguity in the optical timings of BR08 and for his rapid, positive, 
and helpful referee's report. This research has made use of data obtained
from the High Energy Astrophysics Science Archive Research Center (HEASARC),
provided by NASA's Goddard Space Flight Center. Support for this work was
provided in part by NASA through {\it Chandra\/} Award Number GO7-8026X issued
by the {\it Chandra} X-ray Observatory Center, which is operated by the 
Smithsonian Astrophysical Observatory for and on behalf of NASA under contract 
NAS8-03060. NB acknowledges support from NASA contract NAS8-03060 to the
{\it Chandra} X-ray Observatory Center. This work performed under the auspices
of the U.S.\ Department of Energy by Lawrence Livermore National Laboratory 
under Contract DE-AC52-07NA27344.

\references
%-------------------------------------------------------------------------------

% optical data: Table 4 1 value (weighted mean of BVR):
Belle, K.E., et al.,
% Howell, S.B., Mukai, K., Szkody, P., Nishikida, K.,
% Ciardi, D.R., Fried, R.E., \& Oliver, J.P.,
2005, {\it AJ}, {\bf 129}, 1985
\BIBCODE{2005AJ....129.1985B}

% 2nd EUVE observation:
Belle, K.E., Howell, S.B., Sirk, M.M., \& Huber, M.E.,
2002, {\it ApJ}, {\bf 577}, 359
\BIBCODE{2002ApJ...577..359B}

% optical data: Table 3 (3 values)
Beuermann, K., \& Reinsch, K.,
2008, {\it A\&A}, {\bf 480}, 199 (BR08) 
\BIBCODE{2008A&A...480..199B} 

% optical data: Table 2 (40 values):
Bond, I.A., \& Freeth, R.V.,
1988, {\it MNRAS}, {\bf 232}, 753
\BIBCODE{1988MNRAS.232..753B}

% X-ray data: page 450 (1 value):
C\'ordova, F.A., Mason, K.O., \& Kahn, S.M.,
1985, {\it MNRAS}, {\bf 212}, 447
\BIBCODE{1985MNRAS.212..447C}

% optical data: Table 4 (16 values):
Gilliland, R.L.,
1982, {\it ApJ}, {\bf 258}, 576
\BIBCODE{1982ApJ...258..576G}

% optical data: Table 1 (17 values):
Hellier, C., \& Sproats, L.N.,
1992, {\it IBVS}, No.\ 3724 (HS92)
\BIBCODE{1992IBVS.3724....1H}

% optical data: Table 2 (13 values):
Hill, K.M., \& Watson, R.D.,
1984, {\it Proc.\ ASA}, {\bf 5}, 532
\BIBCODE{1984PASAu...5..532H}

% 1st Chandra observation:
Hoogerwerf, R., Brickhouse, N.S., \& Mauche, C.W.,
2004, {\it ApJ}, {\bf 610}, 411
\BIBCODE{2004ApJ...610..411H}

% X-ray data: page L82 (1 value, revised by KM):
Ishida, M., Mukai, K., \& Osborne, J.P.,
1994, {\it PASJ}, {\bf 46}, L81
\BIBCODE{1994PASJ...46L..81I}

% optical data: Table 1 (55 values):
Jablonski, F., \& Busko, I.C.,
1985, {\it MNRAS}, {\bf 214}, 219
\BIBCODE{1985MNRAS.214..219J}

% 2nd Chandra observation:
Luna, G., Brickhouse, N., \& Mauche, C.,
2008, {\it HEAD}, {\bf 10}, \#13.09
\BIBCODE{2008HEAD...10.1309L}

% 1st EUVE observation:
Mauche, C.W.,
1999, {\it ApJ}, {\bf 520}, 822
\BIBCODE{1999ApJ...520..822M}

% RXTE data:
Mukai, K., Ishida, M., Osborne, J., Rosen, S., \& Stavroyiannopoulos, D.,
1998, in Wild Stars in the Old West,
ed.\ S.~Howell, E.~Kuulkers, and C.~Woodward
(San Francisco: ASP), p.~554
\BIBCODE{1998ASPC..137..554M}

% X-ray data: page 554, Table 1 (17 values):
Rosen, S.R., Mason, K.O., \& C\'ordova, F.A.,
1988, {\it MNRAS}, {\bf 231}, 549
\BIBCODE{1988MNRAS.231..549R}

% X-ray data: page 419, Table 1 (6 values):
Rosen, S.R., Mason,  K.O., Mukai, K., \& Williams, O.R.,
1991, {\it MNRAS}, {\bf 249}, 417
\BIBCODE{1991MNRAS.249..417R}

% optical data: Table 3 (89 values):
Sterken, C., et al.,
% Vogt, N., Freeth, R., Kennedy, H.D., Page, A.A., Marino, B.F., \&
% Walker, W.S.G.,
1983, {\it A\&A}, {\bf 118}, 325
\BIBCODE{1983A&A...118..325S}

% optical data: Table 4 (124 values):
Vogt, N., Krzeminski, W., \& Sterken, C.,
1980, {\it A\&A}, {\bf 85}, 106 (VKS80)
\BIBCODE{1980A&A....85..106V}

% optical data: Table 2 (38 values):
Walker, W.S.G., \&  Allen, W.H.,
2000, {\it Southern Skies}, {\bf 39}, 29
\BIBCODE{2000SouSt..39...29W}

%-------------------------------------------------------------------------------
\endreferences

\clearpage

%-------------------------------------------------------------------------------
\begin{center}
Table 1. Times and cycles of spin maxima and $O-C$ residuals.
\begin{tabular}{lccccc}
\hline
& &
\multispan{3}{%
\leaders\hrule height3.3pt depth-3pt \hfill \ $O-C$ (days)\
\leaders\hrule height3.3pt depth-3pt \hfill}& \\
$\rm BJD(TT)-2400000$ &        Cycle  &   VKS80  & HS92       & 
Eqn.~1  & Ref.$^1$\\
\hline
$45546.4450\pn\pm 0.0010\pn$ & 168575 & $-0.013$ & $-0.00090$ & $-0.00039$ & 1\\
$46261.3044\pn\pm 0.0026\pn$ & 183933 & $-0.014$ & $+0.00160$ & $+0.00181$ & 2\\
$46261.3471\pn\pm 0.0025\pn$ & 183934 & $-0.017$ & $-0.00225$ & $-0.00204$ & 2\\
$46261.3928\pn\pm 0.0021\pn$ & 183935 & $-0.018$ & $-0.00310$ & $-0.00289$ & 2\\
$46261.4450\pn\pm 0.0014\pn$ & 183936 & $-0.013$ & $+0.00256$ & $+0.00277$ & 2\\
$46261.4905\pn\pm 0.0017\pn$ & 183937 & $-0.014$ & $+0.00151$ & $+0.00172$ & 2\\
$46261.5353\pn\pm 0.0029\pn$ & 183938 & $-0.015$ & $-0.00024$ & $-0.00003$ & 2\\
$46261.5789\pn\pm 0.0019\pn$ & 183939 & $-0.018$ & $-0.00318$ & $-0.00297$ & 2\\
$46261.6239\pn\pm 0.0015\pn$ & 183940 & $-0.020$ & $-0.00473$ & $-0.00452$ & 2\\
$46261.6730\pn\pm 0.0018\pn$ & 183941 & $-0.017$ & $-0.00217$ & $-0.00196$ & 2\\
$46261.7218\pn\pm 0.0022\pn$ & 183942 & $-0.015$ & $+0.00008$ & $+0.00029$ & 2\\
$46261.7636\pn\pm 0.0014\pn$ & 183943 & $-0.020$ & $-0.00467$ & $-0.00446$ & 2\\
$46261.8148\pn\pm 0.0016\pn$ & 183944 & $-0.015$ & $-0.00001$ & $+0.00020$ & 2\\
$46262.3707\pn\pm 0.0018\pn$ & 183956 & $-0.018$ & $-0.00267$ & $-0.00246$ & 2\\
$46262.4227\pn\pm 0.0017\pn$ & 183957 & $-0.012$ & $+0.00279$ & $+0.00300$ & 2\\
$46262.4668\pn\pm 0.0014\pn$ & 183958 & $-0.015$ & $+0.00034$ & $+0.00055$ & 2\\
$46262.5130\pn\pm 0.0016\pn$ & 183959 & $-0.015$ & $-0.00001$ & $+0.00020$ & 2\\
$46262.5552\pn\pm 0.0020\pn$ & 183960 & $-0.020$ & $-0.00435$ & $-0.00414$ & 2\\
$47328.79044  \pm 0.00154  $ & 206867 & $-0.024$ & $-0.00279$ & $-0.00319$ & 3\\
$47328.88757  \pm 0.00322  $ & 206869 & $-0.020$ & $+0.00125$ & $+0.00084$ & 3\\
$47328.98132  \pm 0.00253  $ & 206871 & $-0.019$ & $+0.00190$ & $+0.00150$ & 3\\
$47329.02481  \pm 0.00155  $ & 206872 & $-0.022$ & $-0.00115$ & $-0.00156$ & 3\\
$47329.16375  \pm 0.00097  $ & 206875 & $-0.023$ & $-0.00185$ & $-0.00225$ & 3\\
$47329.30569  \pm 0.00149  $ & 206878 & $-0.021$ & $+0.00045$ & $+0.00005$ & 3\\
$49185.47425  \pm 0.00023  $ & 246756 & $-0.031$ & $+0.00182$ & $-0.00014$ & 4\\
$49502.17402  \pm 0.00010  $ & 253560 & $-0.033$ & $+0.00186$ & $-0.00043$ & 5\\
$50193.99031  \pm 0.00019  $ & 268423 & $-0.037$ & $+0.00358$ & $+0.00051$ & 6\\
$51683.27876  \pm 0.00010  $ & 300419 & $-0.049$ & $+0.00447$ & $-0.00067$ & 7\\
$51687.51537  \pm 0.00005  $ & 300510 & $-0.048$ & $+0.00539$ & $+0.00025$ & 5\\
$52364.8102                $ & 315061 & $-0.050$ & $+0.00908$ & $+0.00283$ & 8\\
$52364.8608                $ & 315062 & $-0.046$ & $+0.01314$ & $+0.00688$ & 8\\
$52366.7276                $ & 315102 & $-0.041$ & $+0.01810$ & $+0.01184$ & 8\\
$52366.7759                $ & 315103 & $-0.040$ & $+0.01985$ & $+0.01359$ & 8\\
$53027.7678                $ & 329304 & $-0.054$ & $ 0.01206$ & $+0.00462$ & 9\\
$53027.8117                $ & 329305 & $-0.056$ & $ 0.00942$ & $+0.00197$ & 9\\
$53030.8389                $ & 329370 & $-0.055$ & $ 0.01113$ & $+0.00368$ & 9\\
$54235.21476  \pm 0.00007  $ & 355245 & $-0.068$ & $+0.01024$ & $+0.00035$ & 7\\
$54237.96000  \pm 0.00011  $ & 355304 & $-0.069$ & $+0.00927$ & $-0.00063$ & 7\\
$54240.38080  \pm 0.00006  $ & 355356 & $-0.069$ & $+0.00968$ & $-0.00022$ & 7\\
$54243.03427  \pm 0.00007  $ & 355413 & $-0.068$ & $+0.01004$ & $+0.00013$ & 7\\
$54301.68235  \pm 0.00045  $ & 356673 & $-0.069$ & $+0.01023$ & $+0.00019$ &10\\
\hline
\end{tabular}
\end{center}
$^1$References:
1: {\it EXOSAT\/}  (C\'ordova, Mason, \& Kahn 1985),
2: {\it EXOSAT\/}  (Rosen, Mason, \& C\'ordova 1988),
3: {\it Ginga\/}   (Rosen et al.\ 1991),
4: {\it ASCA\/}    (Sequence 20020000),
5: {\it EUVE\/}    (Program IDs 93-067 and 99-009),
6: {\it RXTE\/}    (ObsIDs 10032-01-01 through 10032-01-12),
7: {\it Chandra\/} (ObsIDs 1706 and 7449--7452),
8: optical (ESO La Silla),
9: optical (Beuermann \& Reinsch 2008),
10: {\it Suzaku\/}  (ObsID 402001010).
%-------------------------------------------------------------------------------

\clearpage

%-------------------------------------------------------------------------------
\begin{center}
Table 2. Spin ephemeris constants: $T_{\rm max}=\sum C_nE^n$.
\vskip 3mm
\begin{tabular}{lrrrr}
\hline
Data Included&
\hbox to 0.9in{\hfill $C_0-2400000$\hfill}&
\hbox to 0.9in{\hfill $C_1$\hfill}&
\hbox to 0.9in{\hfill $C_2$\hfill}&
\hbox to 0.9in{\hfill $C_3$\hfill}\\
\hline
\hbox to 1.5in{Optical\leaders\hbox to 0.5em{\hss.\hss}\hfill}
    &  37699.89157  & $   +0.046546478$ & $   -6.25\times 10^{-13}$ &
    \hbox to 0.9in{\hfill $\cdots$\hfill}\\
    & $\pm 0.00054$ & $\pm 0.000000007$ & $\pm 0.22\times 10^{-13}$ &\\
\hbox to 1.5in{EUV \& X-ray\leaders\hbox to 0.5em{\hss.\hss}\hfill}
    &  37699.88930  & $   +0.046546477$ & $   -6.19\times 10^{-13}$ &
    \hbox to 0.9in{\hfill $\cdots$\hfill}\\
    & $\pm 0.00165$ & $\pm 0.000000011$ & $\pm 0.17\times 10^{-13}$ &\\
\hbox to 1.5in{All\leaders\hbox to 0.5em{\hss.\hss}\hfill}
    &  37699.89300  & $   +0.046546454$ & $   -5.85\times 10^{-13}$ &
    \hbox to 0.9in{\hfill $\cdots$\hfill}\\
    & $\pm 0.00041$ & $\pm 0.000000003$ & $\pm 0.05\times 10^{-13}$ &\\
\hbox to 1.5in{All\leaders\hbox to 0.5em{\hss.\hss}\hfill}
    &  37699.89165  & $   +0.046546484$ & $   -7.34\times 10^{-13}$ &
                                          $   +2.16\times 10^{-19}$  \\
    & $\pm 0.00056$ & $\pm 0.000000009$ & $\pm 0.42\times 10^{-13}$ &
                                          $\pm 0.61\times 10^{-19}$  \\
\hline
\end{tabular}
\end{center}
%-------------------------------------------------------------------------------

\vskip 1cm

%-------------------------------------------------------------------------------
\IBVSfigKey{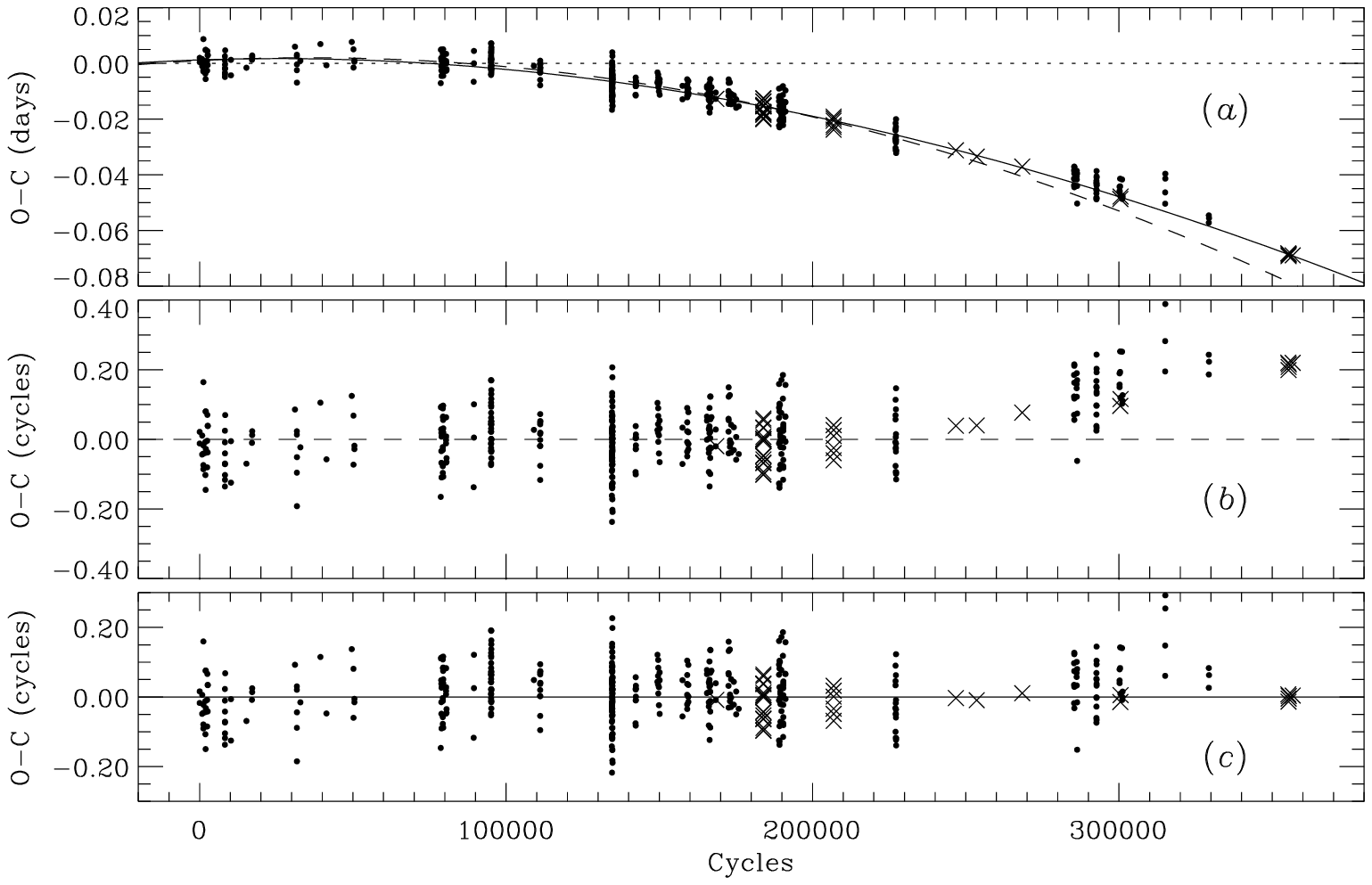}{EX Hya}{O-C diagram}
\IBVSfig{4.0in}{mauche_exhya_fig1.eps}{$O-C$ residuals for the optical
({\it filled circles\/}) and EUV and X-ray ({\it X}s) spin maxima of
EX~Hya relative to
({\it a\/}) the VKS80 linear spin ephemeris,
({\it b\/}) the HS92 quadratic spin ephemeris, and
({\it c\/}) the cubic spin ephemeris of equation 1. In the top panel, 
the HS92 quadratic and equation 1 cubic spin ephemerides are shown 
relative to the VKS80 linear spin ephemeris by the dashed and solid 
curves, respectively.}
%-------------------------------------------------------------------------------

\end{document}